# Developments in Cosmic Growth and Gravitation

ERIC V. LINDER

BERKELEY LAB AND UNIVERSITY OF CALIFORNIA, BERKELEY USA

KOREA ASTRONOMY AND SPACE SCIENCE INSTITUTE, DAEJEON KOREA

**ABSTRACT**

Cosmic surveys of large scale structure have imaged hundreds of millions of galaxies and mapped the 3D positions of over a million. Surveys starting over the next few years will increase these numbers more than tenfold. Simultaneously, developments in extracting information on dark energy, dark matter, neutrinos, and gravity on cosmic scales have advanced greatly, with many important works from Asian institutions.

## 1. INTRODUCTION

The universe is a laboratory where we can probe physics back to the early universe and high energies, uncover properties of particles and fields such as neutrinos, dark matter, and dark energy, and test gravity on scales $10^{15}$ times larger than the solar system.

However, it is also a laboratory where we cannot directly carry out experiments – rather we must observe the photons (and more recently gravitational waves!) that reach us and use well crafted instrumentation, computation, and theory to understand the imprints of physics.

This review article gives a brief overview of several areas on the frontier of understanding the universe through cosmic growth and gravity, focusing on the recent work associated with several Asian institutions and research groups. In §2 we begin at the beginning, discussing how large scale structure grows from initial linear perturbations in density to the rich nonlinear array of the galaxies in the cosmic web. Section 3 goes to the end of time, and reviews how growth will end under the acceleration of cosmic expansion. In §4 we study how growth evolves in the recent past and how new surveys can reveal physics from the cosmic growth history. If we look beyond Einstein's general relativity, then growth changes and observational signatures can test how gravity itself evolves, as outlined in §5. We discuss exciting developments expected in the near future in §6.

## 2. PERTURBING THE UNIVERSE

Large scale structure in the universe starts from small seeds of density perturbations created during early universe inflation and expanded beyond the horizon. As the universe ages and expands, these faint over- and underdensities in matter re-enter the horizon and are released from their frozen state to grow during the matter dominated era.

A small density perturbation $\delta = \delta\rho/\rho$ grows in linear theory from its initial state as $\delta = \delta_i\, D(a)$ with D the growth factor. In the matter dominated era $D \sim a$, where $a$ is the expansion factor of the universe, and so the logarithmic growth rate $f = d\ln\delta/d\ln a = 1$. The growth factor can be measured from the amplitude of the density power spectrum, e.g. from galaxy clustering, and the growth rate from the angular dependence of the anisotropic density power spectrum, since object velocities along the line of sight add or subtract from the cosmic expansion while velocities perpendicular to the line of sight do not. This induced observational anisotropy in a homogeneous density perturbation field is called redshift space distortion (RSD) and is a key probe used by modern galaxy surveys.

Gravitational attraction amplifies density perturbations so eventually the growth in $\delta\rho/\rho$ means that linear theory is not as accurate as needed. Perturbation theory is a main tool for dealing with the quasilinear regime, and then computational simulations once densities become more nonlinear.

Major advances have been made in making perturbation theory more robust, and extending its accuracy further into small scale clustering ($\leq$ 50 Mpc/h). Three main challenges need to be overcome for perturbation theory to be able to extract the desired information from galaxy surveys: 1) the linear matter density perturbations predicted by theory (in "real" space, i.e. the intrinsic rather than observed positions) have to be mapped onto the nonlinear galaxy number density perturbations (with galaxies as biased tracers of density), 2) RSD and other velocity effects have to be taken into account to give the observed ("redshift space") anisotropic power spectrum, and 3) other influences such as massive neutrinos or modified gravity must be allowed for.

Several groups from Asian institutions, often in cross-Asian collaborations, have made major strides in addressing these problems and increasing the accuracy with which data can be interpreted cosmologically. We give a brief selection and description of recent works.

Taruya et al. (2018) present GridSPT, a highly efficient prescription for the use of standard perturbation theory to translate from Gaussian linear fields to quasinonlinear nonGaussian ones. A key aspect is expanding the density field $\delta$ to higher order as

$$\delta(\mathbf{x}) = \Sigma_n e^{n \ln D} \delta_n(\mathbf{x}) \quad (1)$$

and proceeding through a recursion relation for $\delta_n$. They show this is quick and accurate for R > 10 Mpc/h, and can be applied also to higher order correlation functions.

The mapping from real space to observed redshift space for galaxy halo anisotropic clustering is advanced in Zheng et al. (2018), building on numerical simulation work in Chen et al. (2018). They achieve 1-2% accuracy out to Fourier scales k=0.2 h/Mpc. A key ingredient is the use of higher order density-velocity crosscorrelations.

Modeling of RSD is carried forward in Song et al. (2018). This will be crucial for upcoming galaxy surveys such as DESI, and they reach 1% accuracy to k=0.18 h/Mpc. Importantly, they show that scaling methods work to convert results in one cosmological model to another, greatly simplifying calculations. For example, one of the nonlinear density-velocity contributions involves terms like

$$A(k,\mu) = (G_\delta/\hat{G}_\delta)^2 (G_\theta/\hat{G}_\theta) \hat{A}_1 + (G_\delta/\hat{G}_\delta) (G_\theta/\hat{G}_\theta)^2 \hat{A}_2 + \ldots \quad (2)$$

where $G_\delta$ and $G_\theta$ are the density and velocity growth functions, and hats denote the fiducial cosmology where the factors $\hat{A}_1$ and $\hat{A}_2$ are measured from a simulation.

Perturbation theory kernels – the building blocks of clustering statistics – are clearly laid out in Taruya (2016) and generalized for modified gravity including scale dependent screening mechanisms. This is essential for robust tests of cosmic gravity by next generation clustering data. The incorporation of massive neutrinos – which tend to suppress growth – is treated in the $G_\delta$, $G_\theta$ formalism in Oh & Song (2017), improving the use of galaxy data for constraining neutrino mass.

A key final step is the propagation of the galaxy power spectra from perturbation theory to constraining the cosmological parameters. That is, does a 1% fit to the power spectra guarantee accurate cosmology? Osato et al. (2018) compare a number of perturbation theory and effective field theory methods and find excellent results, considering both tightness of constraints (figure of merit) and robustness (figure of bias). Using the response function calibration of Nishimichi et al. (2017) they show that strong, unbiased constraints are achieved out to k=0.24-0.33 h/Mpc, depending on method.

Such advances, and continuing work developing perturbation theory, simulations, and cosmic emulators, add substantial power to the interpretation of data coming from next generation surveys as we seek to understand the growth of cosmic structure, and the role of neutrinos and modified gravity in our universe.

## 3. GRAVITY VS ACCELERATION

Gravity naturally pulls mass together, with more mass having a greater attractive power, leading to gravitational instability. In Newtonian physics this gives an exponential growth from a small overdensity to a massive structure. However, in the expanding universe this is tamed to a power law growth in time. Thus the growth rate reflects a competition between the gravitational force and the cosmic expansion, with growth slowed down but still proceeding forever. A useful way to think of this is trying to join your group of friends by running down the up escalator – despite the social attraction, the stretching of the distance in between makes it difficult to build a larger group.

This suppression of the growth rate is an important cosmological probe of dark energy and we discuss it in the next section. Here we concentrate on the end result of this battle: growth in the far future. During a period of cosmic acceleration of the expansion, the expansion can overwhelm the gravitational attraction and eventually shut off growth. In general relativity the expansion history and growth history are tied together. If we look to theories beyond general relativity, then if gravity strengthens sufficiently over time, could it overwhelm the stretching due to expansion and keep growth going?

For a linear density perturbation $\delta = \delta\rho/\rho$ the growth rate $f = d \ln \delta / d \ln a$ for a subhorizon Fourier mode is given by

$$f' + f^2 + [2 + (\ln H^2)']f - (3/2) G_{eff}(a,k) \Omega_m(a) = 0 \qquad (3)$$

where $a$ is the cosmic scale factor ($a=1$ today), $H = d \ln a/dt$ is the Hubble expansion rate, $\Omega_m(a)$ is the matter density as a fraction of the critical density, and prime denotes $d/d \ln a$. The quantity $G_{eff}$ is the strength of gravity relative to general relativity.

As the universe expands, the matter sourcing the gravitational attraction dilutes with the volume, $\rho \sim a^{-3}$. If the expansion behaves asymptotically as $H^2 \sim a^{-3(1+w)}$ where $w$ is the equation of state parameter of dark energy (i.e. H freezes to a constant de Sitter state for a cosmological constant with $w=-1$) then $\Omega_m(a) = 8\pi G\rho/3H^2$ in the source term vanishes as $a^{3w}$. Thus it becomes negligible and the solution for the late time growth rate is

$$f_\infty = c_f\, a^{(3w-1)/2} \qquad (4)$$

where $c_f$ is a constant (~0.989 for the cosmological constant case $w=-1$). Thus the growth rate goes to zero and growth asymptotically freezes (f dies as $a^{-2}$ as a gets exponentially large in the cosmological constant case). New structures are not formed because matter clumps cannot find each other in the rapidly expanding universe.

Now suppose gravity is modified such that it strengthens with time. Let us take $G_{eff} \sim a^p$ asymptotically. How strong would it need to be to allow the source term to have enough impact to overcome the expansion and prevent the freezing of growth? Since $\Omega_m(a) \sim a^{3w}$ and we want to disrupt a term like $f \sim a^{(3w-1)/2}$ this requires $p > (-3w-1)/2$. This indeed changes the asymptotic behavior of the growth rate to

$$f_\infty = c_f\, a^{p+3w} \qquad (5)$$

but doesn't stop the growth rate from vanishing. Instead we need the source term not to vanish but always have enough effective gravitational attraction to drive growth. This condition is $p + 3w \geq 0$.

These results are shown in Fig. 1. The general relativity (GR) curve is shown in black, and illustrates the rapid decline in the growth rate around today, and its vanishing in the future. The color curves show the results for various strengthenings of gravity $G_{eff} \sim a^p$. Indeed, when $p=3$ then the growth rate does not vanish but goes to a constant, and for $p>3$ growth can increase. (Note that at early times $a \ll 1$, modified gravity with $p<1$ also changes growth at early times – this is generally ruled out by observations, as we discuss in §5.)

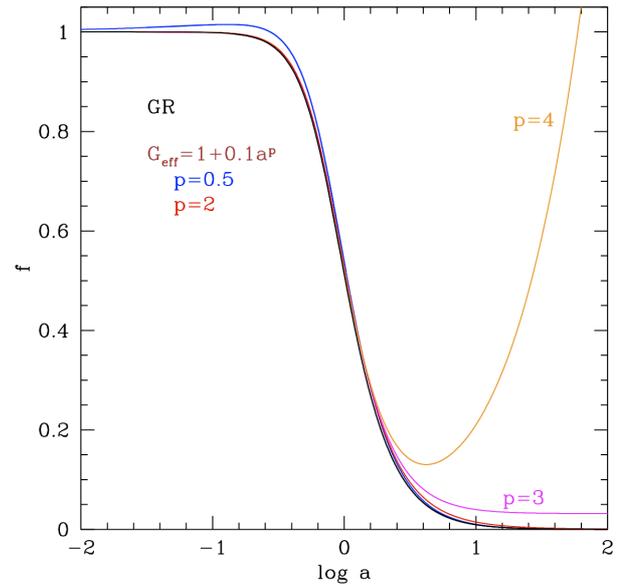

**Fig. 1:** In general relativity (GR), cosmic acceleration suppresses, and eventually shuts off growth. Only modified gravity with sufficiently strengthening gravity can overcome this. Here the background expansion is taken to be as for ΛCDM, the matter plus cosmological constant case (w=-1). [Adapted from Linder & Polarski 2019]

Many theories of modified gravity in the literature have scale dependence in their gravitational force (as noted by the argument k of $G_{eff}$ in Eq. (3)), and hence different Fourier modes of density perturbations grow at different rates. One example of such a theory is f(R) gravity, where the action has an additional term that is a nonlinear function of the Ricci scalar R.

We illustrate the scale dependent growth rate in Fig. 2. To clearly separate the behaviors of the deviation from general relativity, and of one scale from another, we use the gravitational growth index

$\gamma = \ln f / \ln \Omega_m(a)$ (6)

The gravitational growth index is designed to distinguish effects on the growth that arise from sources other than the background expansion, such as modified gravity (Linder 2005, Linder & Cahn 2007). Values of γ that are smaller than in general relativity denote an increased growth rate.

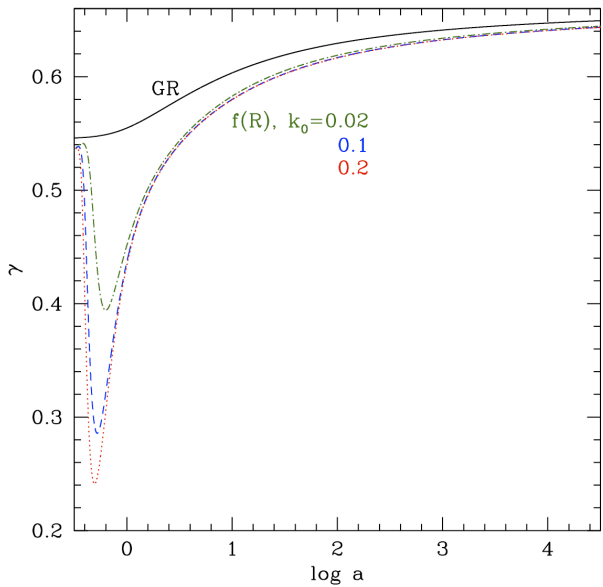

**Fig. 2:** Some modified gravity theories such as f(R) gravity give scale dependent growth, here shown in terms of the gravitational growth index γ for different Fourier modes k in units of h/Mpc. Note however that both the deviation from general relativity (GR) and the scale dependence vanish asymptotically in the future. Growth still halts. [Adapted from Linder & Polarski 2019]

Note that in the recent past there was both maximal deviation from general relativity in γ, and maximal scale dependence. This will be important for the use of cosmic surveys to test gravity through measurements of growth, as we discuss in the next section.

## 4. TESTING ΛCDM WITH COSMIC GROWTH

Many recent surveys, such as the Sloan Digital Sky Survey (SDSS) 3's Baryon Oscillation Spectroscopic Survey (BOSS) and 4's extended BOSS (eBOSS), the Subaru Telescope's FastSound, the Australian-led GAMA and 6dFGRS, and European-led VIPERS surveys, have strived to measure the cosmic growth history. Some upcoming surveys such as the Dark Energy Spectroscopic Instrument (DESI) and the Subaru Measurement of Images and Redshifts (SUMIRE) using the Prime Focus Spectrograph (PFS) are dedicated to this science.

These measurements of the growth rate through studying the anisotropic clustering parallel and perpendicular to the line of sight – redshift space distortions (RSD) – give the quantity $f\sigma_8(a)$, where $\sigma_8$ is a measure of the growth amplitude. One can then compare this to the theory prediction in a particular cosmological model, with a particular matter density, and dark energy or gravity properties. However, it is also of interest to carry out the analysis in a model independent manner, enabling more general tests of the cosmology.

This is the approach taken in Shafieloo et al. (2018), and they test the Friedmann-Lemaître-Roberston-Walker framework and general relativity. They find consistency with the standard cosmology, but note that increasingly accurate measurements over a broader range of expansion and growth history is needed to resolve – or focus in on – currently insufficiently statistically significant tensions.

Recall from Eq. (3) that the growth rate involves the Hubble expansion parameter H(a), the matter density $\Omega_m$, and gravity. The quantity $\sigma_8$ is proportional to the growth amplitude, and is just an integral over f(a) plus a primordial amplitude (or low redshift normalization). To take into account deviation from general relativity one can use $G_{eff}(a,k)$, as we considered in §3 (and will again in

§5). However one can also use the gravitational growth index approach where Eq. (6) is in the equivalent form $f=\Omega_m(a)^\gamma$. Then, by using model independent analysis of the cosmic expansion quantity $\Omega_m(a) = \Omega_m a^{-3}/[H(a)/H_0]^2$, one can constrain γ and test general relativity.

To obtain H(z) in a model independent manner, one uses measured distances, e.g. from standardized candles such as Type Ia supernovae or standardized rulers such as baryon acoustic oscillations (BAO). The expansion rate H(z) is related to the derivative of the distances with redshift $z = a^{-1} - 1$, but taking direct derivatives of noisy data easily leads to spurious results. Two statistically robust techniques are smoothing (L'Huillier & Shafieloo 2017 and references therein) and Gaussian Processes (Rasmussen & Williams 2006).

The analysis of Shafieloo et al. (2018) finds that the distance measurements are consistent with flat ΛCDM cosmology, and that the gravitational growth index is consistent with the general relativity value of γ=0.55. However, the model space is much more open than merely this, with other values giving equally good fits to current data. The bulk of the data was from z<1, and it will be very interesting to map the growth history out to z=2 or beyond with future data and repeat these tests of the cosmological framework.

## 5. MAPPING GRAVITY WITH COSMIC OBSERVATIONS

The cosmic expansion history, whether ΛCDM or with dark energy with some effective equation of state parameter w(z), can be probed by a wide variety of distance measurements. These include the original discovery mechanism of cosmic acceleration – Type Ia supernovae – as well as the baryon acoustic oscillation standard ruler in galaxy clustering, strong lensing time delays, and the cosmic microwave background radiation. In the future, gravitational wave standard sirens may play a role also.

The cosmic growth history can be probed through redshift space distortions in large scale structure clustering and through weak gravitational lensing. More than just the contents of the universe, the growth history can map out the strength of gravity over cosmic time. Just as for distances, growth is an integral relation, depending on cosmic properties over an extended redshift interval. One could parametrize the strength of gravity relative to its value in general relativity, Newton's constant $G_N$, by $G_{eff}(a)$. However, there is no clear time dependence, or even functional form, that is generally applicable to many models, the way there is with the dark energy equation of state, i.e. $w(a)=w_0+w_a(1-a)$, that has been tested to 0.1% precision in the observables (de Putter & Linder 2008).

Thus we seek to find some method for subpercent accurate characterization of gravitational growth suitable for growth observables. For high redshift gravitational modifications, at z≥3 (and recall that many gravity theories, in particular much of the Horndeski class, predict such matter era variations), Denissenya & Linder (2017) showed that one could analytically derive the effect on cosmic growth observations at survey redshifts from modified gravity at high redshifts.

The result was a multiplicative offset in the main growth observable, the logarithmic growth rate $f\sigma_8$ central to redshift space distortions, by a factor proportional to the area under the curve of $G_{eff}(a)/G_N - 1$ over the growth history. That is, one does not have to know the specific, model dependent form $G_{eff}(a)$ but only a single phenomenological parameter corresponding to the integrated deviation, or area, as shown in Eq. (7):

$$\frac{\delta g}{g_\Lambda} \approx \frac{3}{5} \int_0^a d\ln a' \, \delta G_{\text{eff}} \left[1 - \left(\frac{a'}{a}\right)^{5/2}\right]$$
$$\approx \frac{3}{5} \times \text{Area} \qquad (7)$$

where g=D/a is the normalized growth factor.

By comparing to exact solutions for a wide variety of phenomenological modified gravity behaviors for $G_{eff}(a)$ it was found that the area approximation reproduces the growth observables to ≲ 0.3% in the growth factor and ≲ 0.6% in the RSD quantity $f\sigma_8(z \approx 1)$ in most cases, better than the measurement precision of next generation surveys.

The next step was to incorporate gravity modifications that occur at z<3, in particular during the late time, cosmic acceleration epoch (and may well drive the acceleration). Denissenya & Linder (2018) established subpercent accurate fits to the redshift space distortion observable $f\sigma_8(a)$ using two parameters binned in redshift for $G_{eff}(a)$. The results were tested against actual modifications with time dependence that rises, falls, is nonmonotonic, is multipeaked, or corresponds to exact f(R) and braneworld gravity theories (see Fig. 3).

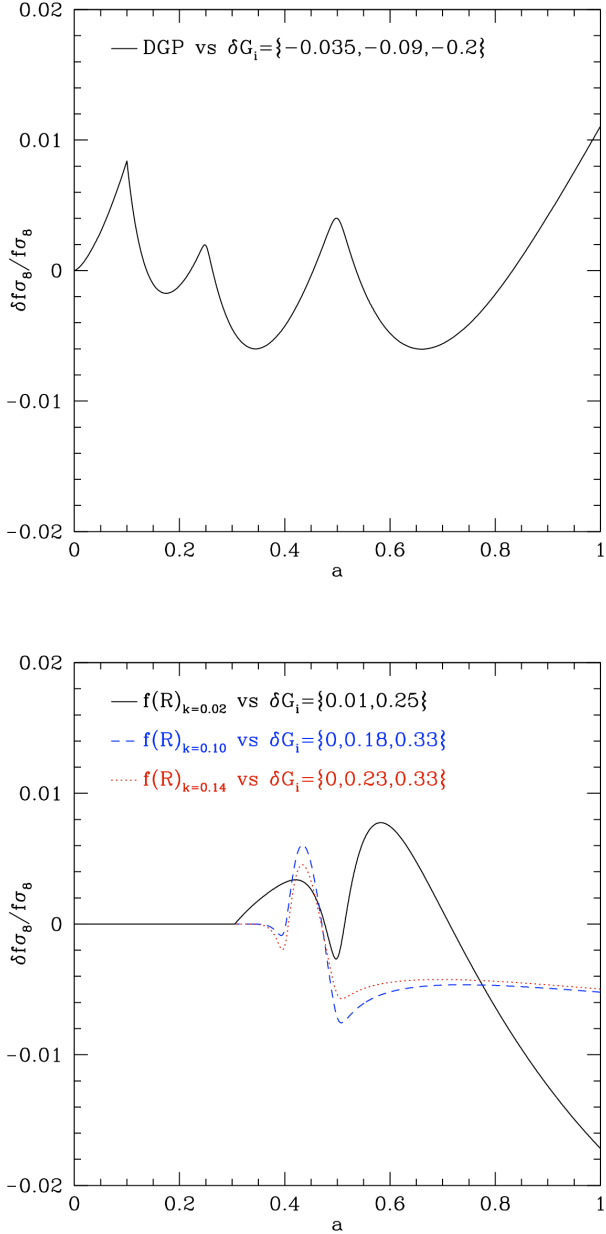

**Fig. 3:** The accuracy of fitting the observational RSD factor $f\sigma_8$ with one early and two late time bins for modified gravity $\delta G_{eff}(a)$ is compared to that for the exact theory cases. We show the fit accuracy for DGP gravity (top panel) and f(R) gravity (bottom panel) with its scale dependent growth (at k = 0.02, 0.1, 0.14 h/Mpc for black, blue, red curves).

The fit residuals were propagated to cosmological parameter biases for DESI observations, and found to cause a shift in the dark energy joint confidence contour by less than the equivalent of ~ 0.1σ. The proposed 2 parameter (3 parameter if data shows the need to incorporate the high redshift "area" parameter) modified gravity description was shown to be not only phenomenologically successful, but to reveal physical characteristics of the underlying theory. Based on the signatures of the bin values, steepness between them, and any need for an early bin or scale dependence, this approach can guide the search for the laws of cosmic gravity in the appropriate direction.

## 6. SUMMARY AND FUTURE DEVELOPMENTS

After an exciting two decades following the discovery of cosmic acceleration learning to map distances and the cosmic expansion history with greater precision, we are now into the age of mapping the cosmic history and gravity with surveys of millions of galaxies in spectroscopy and billions in two dimensional imaging. This can reveal the development of large scale structure from primordial quantum fluctuations, measure the mass of neutrinos, probe the nature of dark energy, and test general relativity – if we are clever enough in our analysis and understanding. I have given a brief overview of some of the challenges, and how they are being met.

New and growing centers of cosmology in the Asia-Pacific region, and dynamic interactions between them have been at the root of many of these developments. Section 2 details work done at CosKASI, the cosmology group at the Korea Astronomy and Space Science Institute (KASI) – awarded First Mover status – in conjunction with groups at the University of Tokyo and Yukawa Institute for Theoretical Institute in Japan, as well as interaction with researchers from Beijing and Shanghai, China. The project in §3 started at the 5[th] Korea-Japan Workshop on Dark Energy and owes much to long term hospitality at KASI. Section 4 is part of a long running research program at KASI, highlighting the central role of statistical techniques in learning about the structure of our universe. The Energetic Cosmos Laboratory is a recently formed center at Nazarbayev University, Kazakhstan, delving deep into the description of cosmic physics outlined in §5, and part of the work was done during an exchange program at KASI.

The research described in this article is just a small part of the active Asian-Pacific growth in cosmology. It is heartening not just to see this development, but the stimulating role that interaction between Asia-Pacific centers – through research collaborations, joint workshops, and exchanges of people – plays in this success.

**Acknowledgement:** I am grateful to my collaborators at many Asian institutions, especially at the Korea Astronomy & Space Science Institute, and the opportunities to participate in many workshops of the China-Japan-Korea groups.

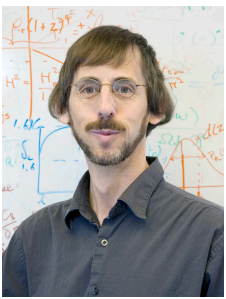

**Eric Linder** is a Research Professor at the University of California, Berkeley and co-founder of the Institute for the Early Universe at Ewha Womans University in Seoul, Korea in 2008 and the Energetic Cosmos Laboratory at Nazarbayev University in Astana, Kazakhstan. He has served as Leadership Scientist at the Korea Astronomy and Space Science Institute and was awarded a Benjamin Lee Professorship at the Asia-Pacific Center for Theoretical Physics. His research field is cosmology, gravity, and particularly the origin and nature of cosmic acceleration.